\documentclass[reprint,aps,prd,superscriptaddress,preprintnumbers,nofootinbib]{revtex4-2}

\usepackage{amsfonts}
\usepackage{amsmath}
\usepackage{amsthm}
\usepackage{amssymb}
\usepackage{array}
\usepackage{dcolumn}

\usepackage{placeins}

\usepackage[svgnames]{xcolor}
\usepackage[hidelinks]{hyperref}
\hypersetup{
    colorlinks=true,
    linkcolor=NavyBlue,
    filecolor=NavyBlue,
    urlcolor=NavyBlue,
    citecolor=NavyBlue,
    }

\usepackage{graphics}
\usepackage{tikz}
\usepackage{graphicx}
\usetikzlibrary{positioning}
\usepackage{caption}
\usepackage{adjustbox}
\usepackage{braket}
\usepackage{enumitem}
\usepackage{ragged2e}
\usepackage{placeins}

\newcommand\be{\begin{equation}}
\newcommand\ee{\end{equation}}
\newcommand\bea{\begin{eqnarray}}
\newcommand\eea{\end{eqnarray}}

\newcommand\gev{\rm GeV}

\begin{document}

\title{Gravitational Waves and Cosmological Observables from First-Order Phase Transitions: Thermal Corrections at Low Temperature}

\author{Katharena Christy}
\email{chri3448@hawaii.edu}
\affiliation{Department of Physics and Astronomy, University~of~Hawaii,  Honolulu, Hawaii~ 96822, USA}

\author{James B. Dent}
\email{jbdent@shsu.edu}
\affiliation{Department of Physics, Sam~ Houston~ State~ University, TX~ 77341, USA}

\author{Sumit Ghosh}
\email{gsumit@manit.ac.in}
\affiliation{Department of Physics, Maulana~Azad~National~Institute~of~Technology, Bhopal~462003, India}

\author{Jason Kumar}
\email{jkumar@hawaii.edu}
\affiliation{Department of Physics and Astronomy, University~of~Hawaii,  Honolulu, Hawaii~ 96822, USA}

\author{J. O'Thello Ward}
\email{juston_ward1@baylor.edu}
\affiliation{Department of Physics, Sam~ Houston~ State~ University, TX~ 77341, USA}
\affiliation{Department of Physics and Astronomy, Baylor University, TX~ 76706, USA}

\begin{abstract}
 We consider the impact on cosmological first-order phase transitions (FOPTs) of low-temperature thermal corrections to the effective potential.  These are corrections from degrees of freedom whose field-dependent masses in the true vacuum are much larger than the nucleation temperature, though in the false vacuum the field-dependent masses  may be much smaller than the nucleation temperature. We focus on the regime in which the thin-wall approximation is valid, and in which these corrections are small enough that they do not alter the vacuum structure of the theory. Although the general form of these corrections to the thermal effective potential can be quite complicated, we argue that the net effect of all such corrections can be well-modeled in this limit with a single new parameter. We determine the shift in the parameters of the FOPT in terms of this new parameter, and the impact on gravitational wave signals and cosmological observables.  
\end{abstract}
\maketitle


\section{Introduction} \label{sec:introduction}

Cosmological first-order phase transitions (FOPTs) in the early Universe can be a source of gravitational waves~\cite{Kosowsky:1992rz,Kosowsky:1992vn,Apreda:2001us,Grojean:2006bp,Huber:2008hg,Weir:2017wfa,Caprini:2015zlo,Mazumdar:2018dfl,Hindmarsh:2020hop,Caldwell:2022qsj,Athron:2023xlk,Croon:2023zay,Caprini:2024hue,Croon:2024mde}. The amplitude and frequency of this signal, if observed, can provide information about fundamental physics parameters of the underlying Lagrangian.  A variety of gravitational wave observatories are operating, or in development, which can potentially observe such signals, and provide a window into fundamental physics in the early Universe, beyond the Standard Model (BSM)~\cite{KAGRA:2021duu,LISA:2024hlh,NANOGrav:2023gor,EPTA:2023fyk,Antoniadis:2022pcn,Xu:2023wog,Abac:2025saz,Evans:2023euw,Sesana:2019vho,Komori:2025pjx,Corbin:2005ny,Weltman:2018zrl,Garcia-Bellido:2021zgu,Bertoldi:2021rqk,MAGIS-100:2021etm,Ackley:2020atn}.  In addition, FOPTs can alter cosmological observables in a manner which can also be probed by astrophysical observatories, for example by releasing dark radiation in the form of latent heat.  However, a key requirement for these types of studies is understanding the effect of thermal corrections to the effective potential. It is these thermal corrections which lead to a universe which is in a false vacuum at early times, when the temperature is high, but transitions to the true vacuum with subsequent cosmological evolution to the current epoch.

There is a large body of work focused on the detailed technical considerations underlying the thermal corrections to any particular fundamental Lagrangian~\cite{Laine:2016hma,Croon:2020cgk,Postma:2020toi,Gould:2021dzl,Schicho:2021gca,Guo:2021qcq,Gould:2021oba,Schicho:2022wty,Lofgren:2023sep,Gould:2023ovu,Lewicki:2024xan,Chala:2024xll,Chakrabortty:2024wto,Navarrete:2025yxy,Ekstedt:2024etx}.  But we adopt a different approach here, by considering the general form of thermal corrections which arise from a broad class of fundamental Lagrangians.  By parameterizing the form of these thermal corrections, we can generally relate the effective potential thermal corrections to the thermal properties of the first-order phase transition, and in turn to the gravitational wave signal.  This will yield a type of `dictionary' relating thermal corrections arising from any type of new physics to changes in the frequency and amplitude of the gravitational wave signal arising from the phase transition.

In previous work~\cite{Dent:2022bcd}, a few of the present authors considered the scenario of a quartic scalar potential with high-temperature thermal corrections arising from degrees of freedom whose effective mass is small compared to the temperature.  However, there will generally be another class of thermal corrections arising from heavy degrees of freedom.  Although these corrections may be Boltzmann-suppressed in the true vacuum, they may not be Boltzmann-suppressed in the false vacuum, thus yielding a nontrivial impact on the phase transition. Moreover, since the fields generating these corrections are heavy in the true vacuum, it may be difficult to know what these degrees of freedom are.  Since one cannot calculate the exact thermal corrections without knowing exactly what heavy degrees of freedom are integrated out, our goal will be to parameterize the form of the low-temperature thermal corrections, and to determine how they generally alter the gravitational wave signal and other cosmological observables arising from the phase transition. Indeed, we will find that the corrections to the thermal parameters of the phase transition can be simply parameterized in terms of the uncorrected thermal parameters and a single additional parameter controlling the size of the low-temperature corrections.

In this approach, we are essentially treating the low-temperature thermal corrections as yielding a small perturbation to the phase transition generated by the usual high-temperature thermal corrections.  As such, it is necessary that these corrections be small, and in particular, small enough that they do not alter the vacuum structure of the theory.  Of course, it is possible for the low-temperature corrections to be large enough to alter the vacuum structure, and even change whether there is a phase transition at all.  But such a situation is not amenable to the generalized analysis which we conduct, and would require a detailed study of the complete Lagrangian.

The plan of this paper is as follows.  In Section~\ref{sec:LowTCorrections}, we describe the general form of the thermal corrections in the low temperature limit, and obtain a simple parameterization.  In Section~\ref{sec:Analytical}, we obtain an analytic approximation describing the effect of these corrections on the thermal parameters of the first order phase transition.  In Section~\ref{sec:NumericalAnalysis}, we validate this analytic approximation with a numerical analysis of a quartic potential, using the \texttt{CosmoTransitions} package~\cite{Wainwright:2011kj}. We conclude in Section~\ref{sec:Conclusions}.


\section{Low-Temperature Thermal Corrections to the Effective Potential} \label{sec:LowTCorrections}

We will examine the case of a single scalar field $\phi$ whose perturbative potential is of the form \bea V(\phi,T) = V_0(\phi,0) + V_{CW}(\phi,0) + V_T(\phi,T) \eea where $V_0(\phi,0$) is the zero-temperature tree-level potential, $V_{CW}(\phi,0)$ is the one-loop zero-temperature potential of the Coleman-Weinberg form, and $V_T(\phi,T)$ is the thermal part of the one-loop potential.  For later convenience, we will write the potential in the form $V (\phi . T) = \Lambda^4 \tilde V (\phi / v, T/v )$, where $v = \langle \phi \rangle_{T=0}$, and $\Lambda$ is a constant which sets the energy scale of the thermal effective potential.

We would like to construct thermal corrections to the effective potential that accurately capture the physics of both the high-temperature and low-temperature regimes.  Consider a degree of freedom, $\eta$ (bosonic or fermionic), which will be integrated out.  The expression for the corresponding contribution to the thermal effective potential for $\phi$ is given by (see, for example,~\cite{Quiros:1999jp,Croon:2020cgk}) \bea V_T(\phi,T) = - T \int \frac{d^3 p}{(2\pi)^3} \ln (1 \pm n_{B,F}(E_p , T)) ,  \label{eq:corr} \eea where the plus (minus) sign above is used for a boson (fermion),  $n_{B,F} (E_p,T) = \left[\exp(E_p/T) \mp 1 \right]^{-1}$, $E_p = \sqrt{p^2 + m^2 (\phi)}$, and $m(\phi)$ is the field-dependent effective mass of $\eta$ (for example, if $\eta$ is a real boson, then $m^2 (\phi) = \partial^2 {\cal L}/\partial \eta^2$ evaluated at $\eta = 0$.).

The contribution to the thermal effective potential from a single degree of freedom can be approximated in the high-temperature limit $x \equiv m(\phi)/T \ll 1$ as~\cite{Morrissey:2012db}
 \begin{widetext}
     \bea \label{eq:highT}
\Delta V_B (\phi , T) &=& \frac{T^4}{2\pi^2} 
\left[-\frac{\pi^4}{45} + \frac{\pi^2}{12} x^2 - \frac{\pi}{6} x^3 
- \frac{1}{32} x^4 \ln (x^2 / a_b) + .... \right] ,
\nonumber\\
\Delta V_F (\phi , T) &=& -\frac{T^4}{2\pi^2} 
\left[-\frac{7\pi^4}{360} + \frac{\pi^2}{24} x^2  
+ \frac{1}{32} x^4 \ln (x^2 / a_f) + .... \right] ,
\label{eq:highTcorr}
\eea 
 \end{widetext} 
($\ln (a_b) \sim 5.4076$, $\ln (a_f) \sim 2.6351$) and can be given in the low-temperature limit ($x \gg 1$) as~\cite{Morrissey:2012db}
{\small \bea
\Delta V_{B,F} (\phi , T) &=& 
\pm \left(\frac{x}{2\pi} \right)^{3/2} e^{-x} 
\left(1 + \frac{15}{8x}  + {\cal O}(x^{-2} )\right) .
\label{eq:lowTcorr}
\eea}

We take the vacuum at high temperature to be at $\phi = 0$. In typical examples, one finds there is a symmetry  which is preserved in the high-$T$ vacuum (when $\phi=0$), but which is broken in the low-$T$ vacuum in which $\phi = \phi_+ > 0$. The field-dependent contributions to the mass are then typically $\propto \phi$, since  the fields which contribute typically are massless in the symmetry-preserving vacuum.  We may then take $x \propto \phi / T$.

If one integrates out a degree of freedom which satisfies $m(\phi) \ll T_N$ (where $T_N$ is the nucleation temperature) for all values of $\phi$ in the range between the true and false vacuum, then we will refer to the associated corrections to the thermal effective potential as being in the high-$T$ limit.  These corrections are well-approximated by the expression in eq.~\ref{eq:highTcorr} for temperatures near the nucleation temperature.

There may be additional thermal corrections from degrees of freedom which are not in the high-$T$ limit. If $x \gg 1$, then these thermal corrections are Boltzmann-suppressed (see eq.~\ref{eq:lowTcorr}), and can be ignored.  This is consistent with the intuition that integrating out very heavy fields will not have a dramatic impact on the low-energy effective potential.  If $m(\phi)/ T_N \gg 1$ for all values of $\phi$ between the true and false vacuum, then the thermal corrections associated with this field will not significantly affect the phase transition. However, since $\phi$ is small in the false vacuum, and $m (\phi) \propto \phi$, these thermal corrections may not be negligible for values of $\phi$ near the false vacuum.

We will then focus on the case in which, at small $\phi$, we have $m(\phi) \ll T_N$, while at the symmetry-breaking vacuum we have $m(\phi) \gg T_N$.  In general, to find the exact form of the corrections (at the nucleation temperature) throughout the barrier between the false and true vacuum, one must evaluate eq.~\ref{eq:corr} numerically using the field-dependent masses, yielding a result specific to the model, and not amenable to a generalized analysis. But near the false vacuum, at small $x$, these corrections can be expanded in the high-$T$ limit, and  can be approximated as $f (\Lambda/v)^4 T^4$, where $f$ is a dimensionless constant which is determined by the number and spin of the degrees of freedom which have been integrated out.  However, at the low-$T$, symmetry-breaking vacuum $\phi_+$, the thermal correction essentially vanishes due to the Boltzmann suppression.

Our interest is not in a detailed analysis of any particular model, but a general parameterization of low-temperature thermal corrections, and their effect on the thermal phase transition parameters, and the gravitational wave signal.  As such, we are only interested in the region of parmeter space in which the low-$T$ corrections do not change the qualitative structure of the vacuum.  It is certainly possible for low-$T$ corrections to change the vacuum structure, for example, by shifting the false vacuum to a symmetry-breaking vacuum.  But in this case, a general parametric analysis would not be of much use, as one would need to know the details of the particular Lagrangian in order to understand the vacuum structure.  Instead, our focus is on the case in which small thermal corrections shift the thermal parameters slightly, without changing the qualitative nature of the phase transition, in order to determine how the gravitational wave signature and cosmological observables are  modified for this class of phase transitions.

In light of the considerations above, it is sufficient to model the effect of the low-$T$ thermal corrections as a shift to $V(\phi =0,T)$ by $f(\Lambda/v)^4 T^4$, where $f$ is a dimensionless parameter.


\section{Analytic analysis of the effect on the thermal parameters} \label{sec:Analytical}

The thermal parameters of the phase transition are defined in terms of $\Delta V(T)$ and $S_3 (T)$.  Here, $\Delta V (T) \equiv V(\phi=0,T) - V(\phi=\phi_+,T) >0$, where the false vacuum is at $\phi =0$ and the true vacuum is at $\phi = \phi_+$. $S_3 (T)$ is the 3D Euclidean action, evaluated on the bounce solution to the equation 
\bea \frac{d^2 \phi}{dr^2} + \frac{2}{r} \frac{d\phi}{dr} = - \frac{\partial V (\phi , T)}{\partial \phi} . \eea
In terms of these quantities, the nucleation temperature $T_N$ of the phase transition satisfies the equation 
\bea
\frac{S_3 (T_N)}{T_N} &=& C (T_N),
\label{eq:Scondition}
\eea
where~\cite{Megevand:2016lpr}
\begin{widetext}
 \bea
C(T) &=& 4 \log \frac{T}{H(T)} +\frac{3}{2} \log \frac{S_3 (T)/T}{2\pi} - 4 \log [T d(S_3(T)/T)/dT] + \log[8\pi v_w^3] ,
\label{eq:C}
\eea   
\end{widetext} 
where the Hubble parameter is $H(T)$, and $v_w$ is the wall velocity. $C (T)$ has a logarithmic dependence on the temperature and on the number of relativistic degrees of freedom; for typical models, one finds $C (T =100~\gev) \sim 140$.  For our purposes the relevant point is that $C (T)$ will not change significantly as we vary $f$, which parameterizes the low-temperature thermal corrections. The other thermal parameters are defined as 
\bea
\frac{\beta}{H} &=& \left( T \frac{d(S_3/T)}{dT} \right)_{T_N} ,
\nonumber\\
\xi &=& \left[\frac{g_*}{100} \frac{10 \pi^2}{3} T_N^4 \right]^{-1} 
\left[\Delta V (T_N) - \left. T_N \frac{d\Delta V}{dT}\right|_{T_N}  \right] .
\label{eq:beta_xi_def}
\eea
where $\beta / H$ parameterizes the speed of the phase transition, and $\xi$ is a dimensionless parameterization of the latent heat density (here, we assume that the dark sector and Standard Model (SM) sector have the same temperature, and we generally take $g_* = 100$).

The effect of low-temperature thermal corrections is to shift the thermal effective potential at the false vacuum. This shift to $\Delta V (T)$ will lead to a shift in $S_3 (T)$, which in turn leads to a shift in the nucleation temperature, as well as $\beta/H$ and $\xi$.  To determine the shifts in the thermal parameters, it is necessary to solve for $S_3(T)$.  Generally, this requires a numerical calculation.  However, once $S_3 (T)$ has been determined in the absence of low-temperature corrections, one can analytically estimate the shift to $S_3 (T)$ due to these corrections.  
Note that our analysis does not rely on any particular choice of the tree-level potential.
We only assume that, at high temperatures, the vacuum is symmetry preserving, while symmetry is broken in the low-temperature vacuum.

If we take the thin wall approximation, then we may make the estimate~\cite{Dine:1992wr} 
\bea
S_3 (T) &\sim & -\frac{4\pi}{3}R^3 \Delta V(T) + 4\pi R^2 \Sigma (T) , 
\nonumber\\
\Sigma (T) &\sim& \int_{0}^{\phi_+} d\phi ~\sqrt{2 V(\phi , T)} .
\label{eq:S3}
\eea
where $R$ is the size of a true vacuum bubble which is nucleated at temperature $T$ and $\Sigma$ is the bubble wall surface energy density.  The volume term is the Euclidean Lagrangian integrated over the bubble, while the surface area term arises from integrating the Lagrangian evaluated at the bounce solution across the bubble wall. Note that the expression for $S_3$ in Eq.~\ref{eq:S3} is approximately the free energy density of the true vacuum bubble.  We see that vacuum energy density term will tend to cause the bubble to expand, while the bubble wall energy density term will 
tend to cause the bubble to contract. 

To proceed further, it is necessary to estimate the bubble wall size.  To do this, we can consider a critical bubble which satisfies $\Delta p \propto -dS_3 / dR \sim 0$.  For such a bubble, 
\bea
4\pi R_c^2 \Sigma (T) &\sim& 2\pi R_c^3 \Delta V(T) , 
\nonumber\\
S_3 (T) &\sim& \frac{2\pi}{3}R_c^3 \Delta V(T) \sim \frac{16 \pi}{3} \frac{\Sigma (T)^3}{\Delta V(T)^2} .
\eea
Note that for $R<R_c$, we will have $\Delta p < 0$, and the true vacuum bubble will tend to contract. As such, we are justified in assuming that the nucleated bubble is critical.

The effect of changing $f$ from zero will be to shift $\Delta V$ by $\delta \Delta V(T) = (\Lambda/v)^4 f T^4$.  Since the height of the barrier is assumed to be large compared to $\delta \Delta V$, we may assume that $\Sigma$ changes negligibly as $f$ is varied, and $S_3 \propto \Delta V^{-2}$.  
This amounts to saying that the effect of the low-temperature thermal corrections on the bubble wall energy density is 
negligible, implying that the detailed behavior of the low-temperature thermal corrections at values of $\phi$ between the false and true 
vacua is not important.  Rather, the high-$T$ thermal corrections which dominate the behavior of the barrier arise from fields whose 
field-dependent masses are small compared to the nucleation temperature throughout the region in field-space between the 
false and true vacua.
This fact is what will justify parameterizing the low-$T$ corrections entirely in terms of 
the shift to the height of the false vacuum.

We then find that 
\bea
\frac{\delta S_3 (T)}{T} &\sim& \frac{S_3 (T)}{T} \left[ -\frac{2}{\Delta V (T)} 
\left(\frac{\Lambda}{v} \right)^4 fT^4 \right] ,
\nonumber\\
&\sim&  -\frac{2 C(T) }{\Delta V (T)} 
\left(\frac{\Lambda}{v} \right)^4 fT^4 .
\eea

The shift to $S_3$ will lead to a small shift in the nucleation temperature, while  $C(T)$ varies only by a very small amount.  We thus find
\bea
0 &=&  \delta T_N\left. \frac{d(S_3/T)}{dT} \right|_{T_N} + \frac{\delta S_3 (T_N)}{T_N} .
\eea

We then have
\bea
\frac{\delta T_N}{T_N} &\sim & \frac{2 C(T_N)}{\beta/H }   
\frac{f (\Lambda/v)^4 T_N^4}{\Delta V (T_N)}.
\label{eq:deltaTandf}
\eea
where we have used Eq.~\ref{eq:beta_xi_def}. This result suggests that $T_N$ should increase linearly with $f$.  Moreover, this result does not depend on the bubble wall energy density $\Sigma$.  We thus see that the corrections to the nucleation temperature will not depend on the details of the behavior of the low-temeperature corrections within the barrier, thus justifying our simplified parametric treatment of the low-temperature corrections.

Finally, we see that if $C(T_N) \ll \beta / H$, then $\delta T_N / T_N \ll \delta \Delta V(T_N) / \Delta V(T_N)$.  We will mostly focus on the case in which the transition rate is large, and this limit can be taken.  Note, however, that the thermal parameters will depend on  quantities such as $[d\Delta V(T)/dT]_{T_N}$,  but there is no generic way to relate the variation in these quantities under a shift in $T_N$ to the thermal parameters.  However, if the variation in these quantities were large (that is, if $|\delta T_N [d^2\Delta V/d^2T]_{T_N}| \sim |d\Delta V/dT_N|$), it would suggest that the FOPT was likely supercooled.  To gain approximate understanding of the thermal parameters, we will assume that this is not the case, and the variation in $\delta T_N$ can be ignored. We then find
\bea
\delta \xi &\sim&  \left[\frac{g_*}{100} \frac{10 \pi^2}{3} T_N^4 \right]^{-1} 
\left[\delta \Delta V(T_N) - T_N \frac{d\delta \Delta V(T_N)}{dT_N}   \right]
\nonumber\\
&\sim& -\frac{9}{10\pi^2} f  \left[ \left(\frac{g_*}{100}  \right)^{-1} 
\left(\frac{\Lambda}{v} \right)^4 \right] .
\eea
The shift in the latent heat parameter does not depend on the nucleation temperature.  It is also interesting to note that the sign of the shift may seem somewhat counterintuitive; as $f$ increases, the difference between the thermal effective potential at the false and true vacua increases, but the latent heat released in the transition decreases.  The reason is because the difference in the thermal effective potential essentially amounts to the difference in the free energy density between the false and true vacua, but the difference in the energy density has an entropy term as well.  This entropy term more than compensates for the increased difference in the effective potential. 

Finally, we have (taking $C(T_N) / (\beta / H) \ll 1$)
\begin{widetext}
    \bea
\delta\left( \frac{\beta}{H}\right) &\sim& \left. T \frac{d(\delta S_3 /T)}{dT} \right|_{T_N}  ,
\nonumber\\
&\sim&  -\frac{8 C(T_N) }{\Delta V (T_N )} \left(\frac{\Lambda}{v} \right)^4 fT_N^4 
+ \frac{2 C(T_N) }{\Delta V (T_N )^2} 
\left(\left(\frac{\Lambda}{v} \right)^4 fT_N^4 \right) \frac{d\Delta V(T_N)}{dT_N} T_N ,
\nonumber\\
&\sim& -\frac{ C(T_N) }{\Delta V (T_N )} \left(\frac{\Lambda}{v} \right)^4 fT_N^4 
\left[6 + 2\xi \left[\left(\frac{g_*}{100}\right) \frac{10 \pi^2}{3} \frac{T_N^4}{\Delta V (T_N)} \right] \right] ,
\nonumber\\
\eea
\end{widetext} 
implying that $\beta/H$ decreases roughly linearly with increasing $f$.

Note that $\delta T_N /T_N$ and $\delta (\beta/H)$ are both proportional to $f (\Lambda/v)^4 T_N^4 / \Delta V(T_N)$.  Although the signs of the changes of these thermal parameters are fixed, the magnitudes  depend on the shift to $S_3$, which  in turn depends on the bubble wall size and energy density through equation Eq.~\ref{eq:S3}.  It is here that the detailed dependence on the scalar profile enters.  Taking a critical size bubble and assuming the bubble wall energy density is nearly fixed yields the scaling we have found above.  But $\delta \xi$ has no such dependence on $S_3$, making this estimate more robust. Moreover, we should find
\begin{equation}
\scalebox{0.95}{$\displaystyle\frac{[\delta (\beta/H)] / [\beta/H]}{\delta T_N / T_N} = 
- 3 - \xi \left[\left(\frac{g_*}{100}\right) \frac{10 \pi^2}{3} \frac{T_N^4}{\Delta V (T_N)} \right]$} 
\end{equation}
where the dependence on $f$ has factored out.  Note in particular that the fractional change in $\beta / H$ will be much larger than the fractional change in the nucleation temperature.

\subsection{Corrections to the gravitational wave signal}

The gravitational wave signal produced by an FOPT depends in detail on the thermal parameters of the phase transition~\cite{Bodeker:2009qy,Espinosa:2010hh,Bodeker:2017cim,BarrosoMancha:2020fay,Giese:2020rtr,Balaji:2020yrx,Hoche:2020ysm,Giese:2020znk,Guo:2021qcq,Ai:2021kak,Laurent:2022jrs,Wang:2022lyd,Lewicki:2022nba,Tenkanen:2022tly,Krajewski:2023clt,Cai:2023guc,Wang:2023jto,Wang:2023kux,Dorsch:2023tss,Ekstedt:2024fyq,Tian:2024ysd}.  The GW signal itself can have several components -- sourced by sound waves in the plasma, by turbulence, and by bubble collisions~\cite{Weir:2017wfa,Caprini:2019egz,Gould:2021dpm}.  Although the sound wave contribution is usually dominant~\cite{Hindmarsh:2013xza,Hindmarsh:2015qta}, the determination of the amplitude and spectrum of this signal is an evolving field, which depends in detail on results from numerical simulations of bubble formation and expansion~\cite{Hindmarsh:2016lnk,Cutting:2018tjt,Cutting:2020nla,Hindmarsh:2019phv,Guo:2020grp,Guo:2021qcq,Guo:2024gmu,Wang:2025eee,Giombi:2025tkv}.  However, the rough dependence of the signal on the parameters $\beta / H$ and on $\xi$ is determined by general power-counting considerations, and is generally robust.

For example, the frequency scale of the sound wave signal (which one can think of as the frequency $f_{sw}$ at which the amplitude is maximized) scales as $\propto \beta / H$, since this is the only quantity which is determined by the time-scale of the FOPT.  We thus see that increasing $f$ will correspond to decreasing $f_{sw}$.

The peak amplitude of the sound wave gravitational signal scales as 
\be 
h^2 \Omega_{sw}^{max} \propto \xi^{2n+2} (\beta / H)^{-1}, 
\ee 
where $n$ is a number between $0$ and $1$ determined by the wall velocity, and  reflects the efficiency for converting latent heat into sound waves~\cite{Espinosa:2010hh}.  When the production of sound waves is most efficient (with an efficiency close to $1$), then $n$ tends to be small. We thus see that although the shift in $h^2 \Omega_{sw}$ will be linear in $f$, the sign of the proportionality is determined by the competition between the dependence on $\xi$ and $\beta / H$.  

For concreteness, we choose~\cite{Dent:2022bcd}
\begin{widetext}
    \bea
f_{sw}^{max} &=& 8.9 \times 10^{-3} {\rm mHz} \left( \frac{\beta}{H} \right) \left( \frac{T_N}{100~\gev} \right) 
\left(\frac{g_*}{100} \right)^{1/6} v_w^{-1},
\nonumber\\
h^2 \Omega_{sw}^{max} &=& 8.5 \times 10^{-6} ~ \xi^{2+2n} \left(\frac{\beta}{H} \right)^{-1} 
\left(\frac{g_*}{100} \right)^{1/3} v_w.
\eea
\end{widetext} 

In general, the amplitude and peak frequency of the GW signal depends on the bubble wall velocity $v_w$ (we will simply take as $v_w \rightarrow 1$).  The bubble wall velocity in turn depends on the details of the model, including the number of relativistic degrees of freedom in the true and false vacua (see, for example, Ref.~\cite{Marfatia:2020bcs}).  We do not study this in detail here because, in the relativistic limit, it does not strongly affect the correction to the GW signal due to the low-$T$ thermal corrections.  In the non-relativistic limit, the shift in the latent heat parameter would in turn lead to a non-negligible shift to the bubble wall velocity, which could be determined for any particular model~\cite{Marfatia:2020bcs}, leading to a model-dependent shift to the peak frequency and amplitude of the GW signal~\cite{Dent:2022bcd}.


\section{Numerical Analysis Using a Quartic Potential} \label{sec:NumericalAnalysis}

Thus far, our analytic analysis has been general, not assuming any particular form of the effective potential. But to verify those results with a numerical analysis, we will have to choose a specific example. We will focus on the case in which the tree-level zero-temperature potential is quartic, with a $\phi \rightarrow -\phi$ symmetry:
\bea
V_0(\phi,0) &=& \Lambda^4 \left[-\frac{1}{2} \left(\frac{\phi}{v} \right)^2 + \frac{1}{4} \left(\frac{\phi}{v} \right)^4 \right],
\eea
where the global minimum of the zero-temperature potential is $\phi = \pm v$. For simplicity, we will assume that the zero-temperature Coleman-Weinberg potential $V_{CW}(\phi,0)$ does not qualitatively affect the vacuum structure, and can be ignored.  But we emphasize that our general results do not depend on these assumptions.
 
In the high-$T$ limit, then, the dominant thermal corrections to the effective potential are proportional to $\phi^2 T^2$ and $\phi^3 T$.  The term proportional to $T^4$ is $\phi$-independent, and can be ignored because it does not affect the phase transition. The log term has only a logarithmic dependence on temperature, and can be combined with the zero-temperature Coleman-Weinberg potential~\cite{Quiros:1999jp,Croon:2018erz}. Adding these high-temperature thermal corrections, we can write the thermal effective potential as~\cite{Croon:2018erz}
\begin{widetext}
    \bea
V_{high}(\phi,T) &=& \Lambda^4 \left[\left( -\frac{1}{2} + c \left(\frac{T}{v} \right)^2 \right) \left(\frac{\phi}{v} \right)^2 
+ b \frac{T}{v} \left(\frac{\phi}{v} \right)^3
+ \frac{1}{4} \left(\frac{\phi}{v} \right)^4 \right] ,
\label{eq:Vhigh}
\eea
\end{widetext}  
where the coefficients $b<0$ and $c$ depend on the details of the degrees of freedom which were integrated out (in the high-$T$ limit).  We are interested in the scenario in which, at sufficiently high-temperature, the global minimum is at $\phi = 0$, while at low enough temperature, the global minimum of the potential is at $\phi > 0$.  These conditions are satisfied if $c/b^2 >1$, with the critical temperature for the phase transition being  $T_C = v/\sqrt{2(c-b^2)}$. Note that a first-order phase transition will occur as a result of these high-$T$ thermal corrections, even in the absence 
of heavy fields. A general analysis of gravitational waves from a first-order phase transition, given this general parameterized form of the thermal effective potential, was done in Ref.~\cite{Dent:2022bcd}.

We may now add thermal corrections arising from degrees of freedom for which $m(\phi=0)/T_N \ll 1$, but $m(\phi_+) /T_N \gg 1$.  As we have seen, the shift to the thermal effective potential at $\phi = 0$ can be parameterized as $f (\Lambda / v)^4 T^4$, but there is no generic expression for the thermal corrections within the barrier, as $\phi$ assumes values between $0$ and $\phi_+$. However, it is necessary for the correction to smoothly go to zero as $\phi$ approaches the true vacuum ($\phi_+$), where the thermal corrections are Boltzmann-suppressed.  To model this, we multiply the low-$T$ thermal correction evaluated at $\phi = 0$ by a window function given by \begin{widetext}
    \bea w (\tilde \phi) &=& \frac{1}{\left( 1 + \exp \left[c_1 (\tilde \phi - \delta/2)\right] \right) \left( 1+ \exp \left[-c_1 (\tilde \phi + \delta/2)\right] \right) } , \eea
\end{widetext} 
where $\tilde \phi \equiv \phi / v$.  $\delta$ defines with the width of the window function, and the constant $c_1$ defines the sharpness of the cutoff, with larger values of $c_1$ corresponding to a sharper cutoff to the window function.  The thermal effective potential can then be written as 
\begin{widetext}
    \bea
V_{thermal}(\phi,T) &=& \Lambda^4 \left[ \left(-\frac{1}{2} + c \tilde T^2  \right) \tilde \phi^2 
+ b \tilde T \tilde \phi^3
+ \frac{1}{4} \tilde \phi^4 + f \tilde T^4 w(\tilde \phi) \right] ,
\label{eq:Vthermal}
\eea
\end{widetext} 
where $\tilde T \equiv T / v$. Note that the parameter $f$ includes the effect of low-temperature corrections from all relevant fields, but from a single degree of freedom, one expects a contribution to $f$ of ${\cal O}(0.1)$ (see Eq.~\ref{eq:highT}).

Adopting this simplified analytic form of the low-$T$ thermal corrections will allow us to scan over a variety of models, without having to perform detailed numerical calculations of the thermal effective potential for each case.  But note that this simplified form of the low-$T$ thermal corrections was not used as part of the analytic analysis in the previous section.  Indeed, that analytic analysis used only a parameterization of the thermal corrections at the false vacuum in terms of the parameter $f$, arguing that the precise manner in which the correction interpolated to zero within the barrier between the false and true vacua was not important.  We can test that conclusion by seeing if the results of the analytic analysis are consistent with a numerical analysis using this simplified version of the low-$T$ thermal corrections, even if this simplified form does not match the exact form of the thermal corrections which one would find with any particular microscopic model.  That is to say, if the analytic analysis, taking as input only the value of the low-$T$ thermal corrections at $\phi=0$, reproduces the thermal parameter corrections found using a numerical analysis of the effective potential with a parameterization of the  thermal corrections for $0 \leq \phi \leq \phi_+$, then it must indeed by correct that the form of the low-$T$ correction within the barrier does not significantly impact the thermal parameters. This would imply that little was lost by using this simple parameterization, rather than the exact form of the thermal corrections for any particular microscopic model.

We must make choices of the parameters $c_1$ and $\delta$ of the window function in line with the considerations above.  We choose $\delta$ to be large enough that the thermal correction is roughly constant over the range between  $\phi=0$ and the middle of the barrier.  This ensures that the edge of the window function does not introduce a new minimum.  We also choose $c_1$ to be large enough that the edge of the window function is sharp, and the thermal correction is insignificant near the true vacuum.  Qualitatively, this amounts to shifting the potential by a temperature-dependent but $\phi$-independent constant near the false vacuum, but leaving the potential unaltered near the true vacuum.  Finally, we restrict $f$ so that the shift to the potential at the false vacuum is small compared to the height of the barrier.

\begin{figure}[tbp]
\centering 
\includegraphics[width=0.46\textwidth]{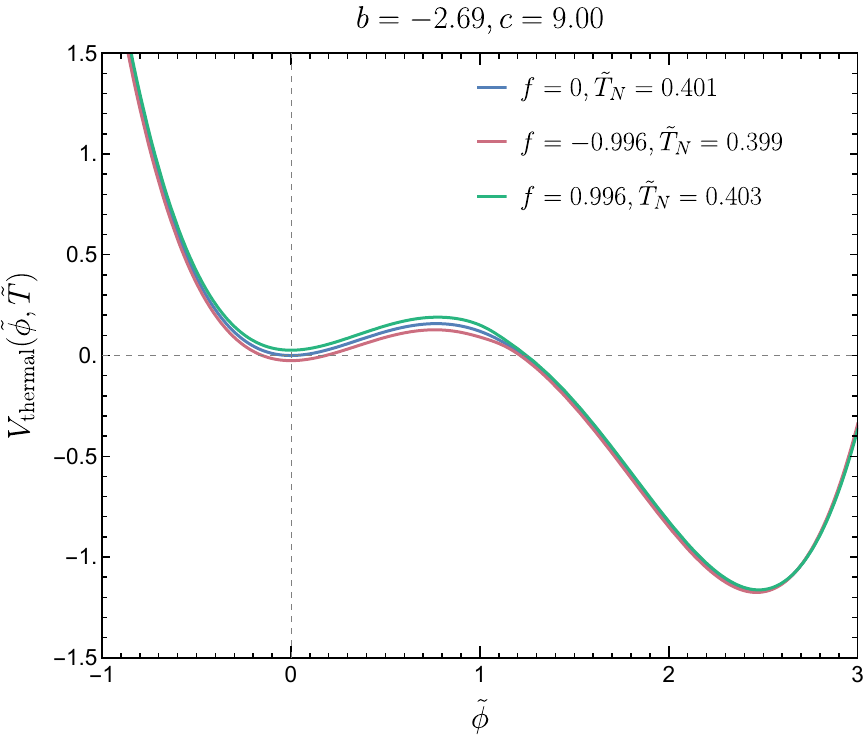}

\captionsetup{justification   = RaggedRight,
             labelfont = bf}
\caption{\label{fig:potential} $V(\tilde \phi, \tilde T_N)$ for the benchmark point 
$(b,c) = (-2.69, 9.00)$, for $f = 0, \pm 0.996$, as labeled.
}
\end{figure}

Specifically, we choose $\delta/2 = \phi_{max}$, where $\delta / 2$ is the positive edge of the window function, and $\phi_{max}$ is the location of the maximum of the potential barrier at $f=0$ and $T=T_C$.  For the case of the quartic potential in Eq.~\ref{eq:Vhigh}, this is given by
\bea
&~&\phi_{max} =  \frac{-3bT_C - \sqrt{9b^2T_C^2 + 4(1-2cT_C^2)}}{2} = -b T_C ,
\nonumber\\
&~&V(\phi_{max}, T_C) = \Lambda^4 \frac{b^4}{16 (c-b^2)^2} .
\eea
To ensure that the low-temperature thermal corrections do not qualitatively change the vacuum structure, we require that, at the critical temperature, the shift of the high-temperature vacuum is less than $1/10$ the height of the barrier (at $T=T_C$).  That is, we require 
\bea
|f| \left(\frac{\Lambda}{v} \right)^4 T_C^4 &<& \frac{1}{10} V (\phi_{max}) .
\eea
We fix $c_1 = 20$.  Choosing such a large value ensures that the window function is relatively constant between $-\delta / 2$ and $\delta / 2$, and thus does not introduce any new minima which might change the vacuum structure.\footnote{In Appendix~\ref{sec:window}, we discuss how our results vary under changes to the window function parameters.}  
 As an illustration, we plot the potential for a benchmark point, for a few choices of $f$, in Figure~\ref{fig:potential}.  Here, the potential is evaluated at the nucleation temperature, which is the temperature at which the bubble nucleation rate exceeds the Hubble expansion rate, as we discuss in the following section.

We have chosen a simple parameterization of the sum of all low-$T$ thermal corrections.  This parameterization will be a good approximation if, for temperatures between the critical and nucleation temperatures, $T \gg m(\phi)$ for $\phi$ near the false vacuum, but $T \ll m(\phi)$ for $\phi$ near the true vacuum.  This parameterization captures well thermal parameters which are evaluated at the true and/or false vacuum, such as $\Delta V$ or $d\Delta V / dT$.  This parameterization does not correctly model in detail the transition in behavior of the thermal effective potential corrections between the high-$T$ and low-$T$ regimes as a function of $\phi$.  This may affect thermal quantities such as the Euclidean bounce solution, but this quantity is generally sensitive to the overall integral of potential barrier, not to its detailed shape.

We emphasize that all thermal corrections need not be correctly modeled by the parametrization which we have chosen.  In particular, thermal corrections which do not obey the assumptions which we have posited above may yield non-trivial phenomena which we do not capture here, including removal of the potential barrier (resulting in a roll-over phase transition, or potentially no transition at all), or a qualitative change in the nature of the true or false vacuum.  Such situations, however, are not amenable to the general parameterized analysis which we perform here.  Instead, one must must determine the thermal effective potential in detail to determine the vacuum structure as a function of $T$.  Our analysis instead focuses on a simple parameterization of a large class of thermal corrections which do not qualitatively change the vacuum structure.

\subsection{Results}

To test how well our analytic analysis performs, we compare those results to the results obtained from a numerical analysis using the \texttt{CosmoTransitions} package~\cite{Wainwright:2011kj}.  

When the low-temperature thermal corrections vanish (that is, $f=0$), we may use an analytic approximation to thermal effective potential presented in Ref.~\cite{Dine:1992wr}.  We then find
\begin{widetext}
    \bea
\left. \frac{S_3 (T)}{T}\right|_{f=0} &=& \frac{4.85 M^3}{E^2 T^3} 
\left[1 + \frac{\alpha}{4} \left(1+ \frac{2.4}{1-\alpha} + \frac{0.26}{(1-\alpha)^2} \right) \right] ,
\label{eq:S3Tapprox}
\eea 
\end{widetext}
where
\bea 
M^2 &=& 2 \frac{\Lambda^2}{v^2} \left(c \frac{T^2}{v^2} - \frac{1}{2} \right)  , \\ 
E &=& -b \frac{\Lambda^4}{v^4}, \\ 
\alpha &=& \frac{M^2 }{2 E^2 T^2} \frac{\Lambda^4}{v^4} . 
\eea

This analytic approximation is generally in good agreement with the numerical results from \texttt{CosmoTransitions}~\cite{Guo:2020grp,Dent:2022bcd}, and can be used as another check of our 
results.\footnote{In Appendix~\ref{sec:BubbleSolution}, we plot the bubble solution 
obtained by \texttt{CosmoTransitions} for one of our benchmark points, for a few values of $f$.}

We will use $v$, the vev of $\phi$ at zero temperature, to parametrize the the energy scale; all other dimensionful quantities ($\Lambda$, $T$ and $\phi$) will be expressed as dimensionless ratios with respect to this scale.  We will also set $C(T) = 140$.  In addition to $v$, $C(T)$ in general depends on the Planck scale, which appears in the Hubble parameter, upon which $C(T)$ depends. The choice $C(T) = 140$ is thus equivalent to a choice of the scale $v$ such that this condition is satisfied.  For typical choices of $g_*$ and the potential, this corresponds to $T_N \sim {\cal O}(100~\gev)$, which can be achieved by a suitable choice of $v$. We will take $g_* =100$ and $\Lambda / v =1$ for all subsequent numerical results. For the purpose of illustrating the applicability of our analytic analysis, these choices will be sufficient. For a very different choice of $v$, $C(T)$ can be found by solving Eq.~\ref{eq:Scondition} using Eq.~\ref{eq:C}.  

\begin{widetext}
    
\begin{figure}
    \centering
    \includegraphics[width=0.95\linewidth]{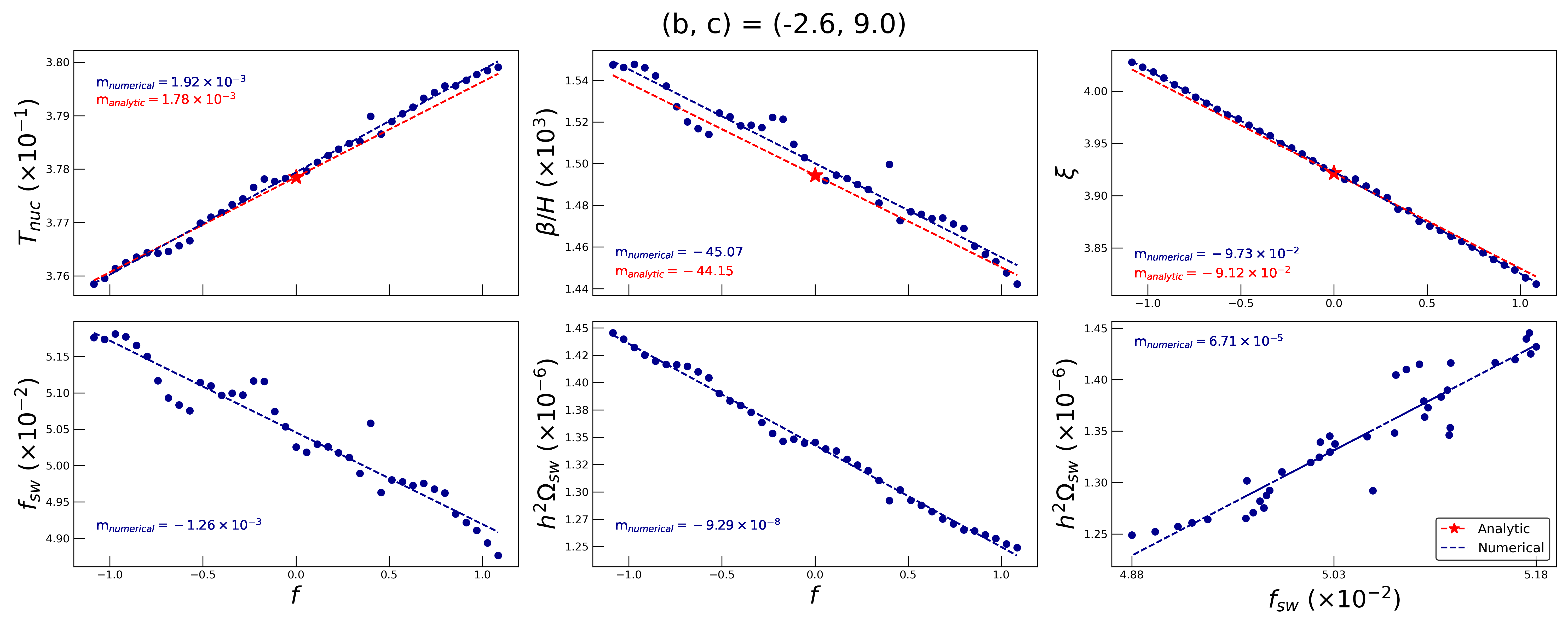} \\ 
    \includegraphics[width=0.95\linewidth]{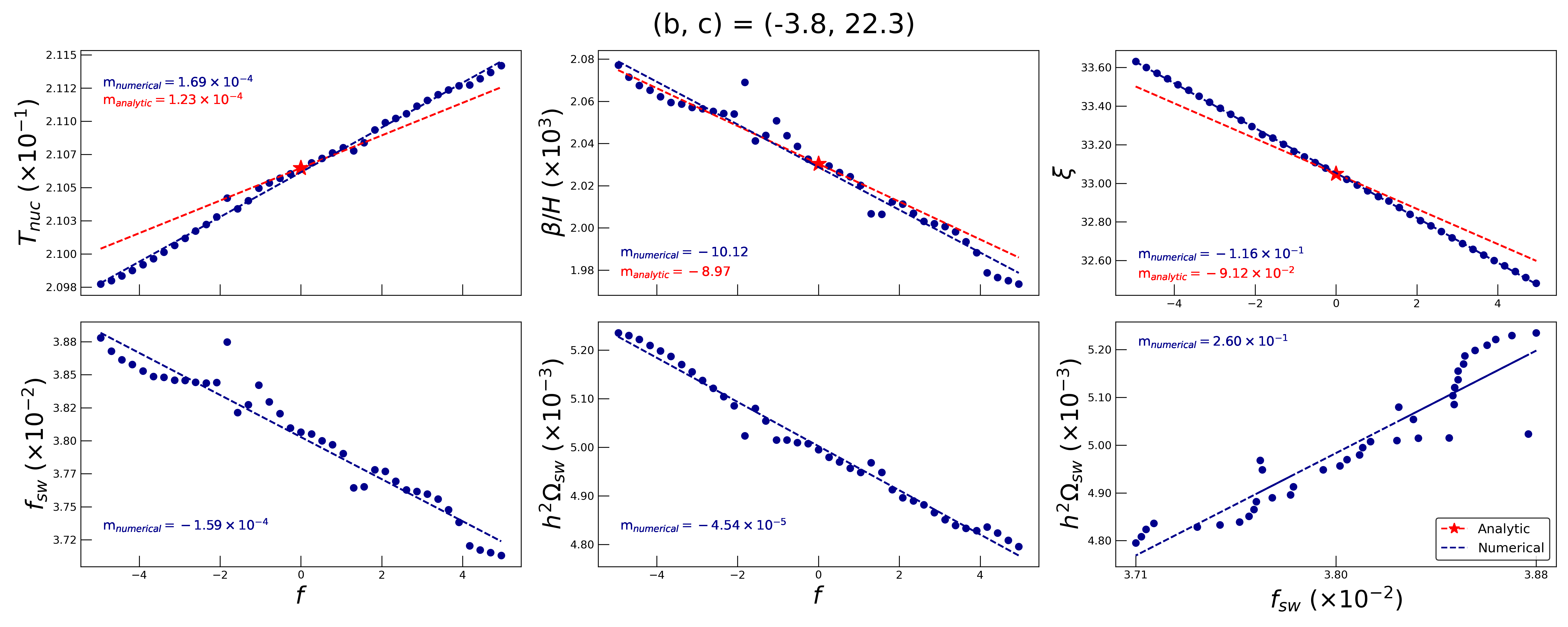} \\
     \includegraphics[width=0.95\linewidth]{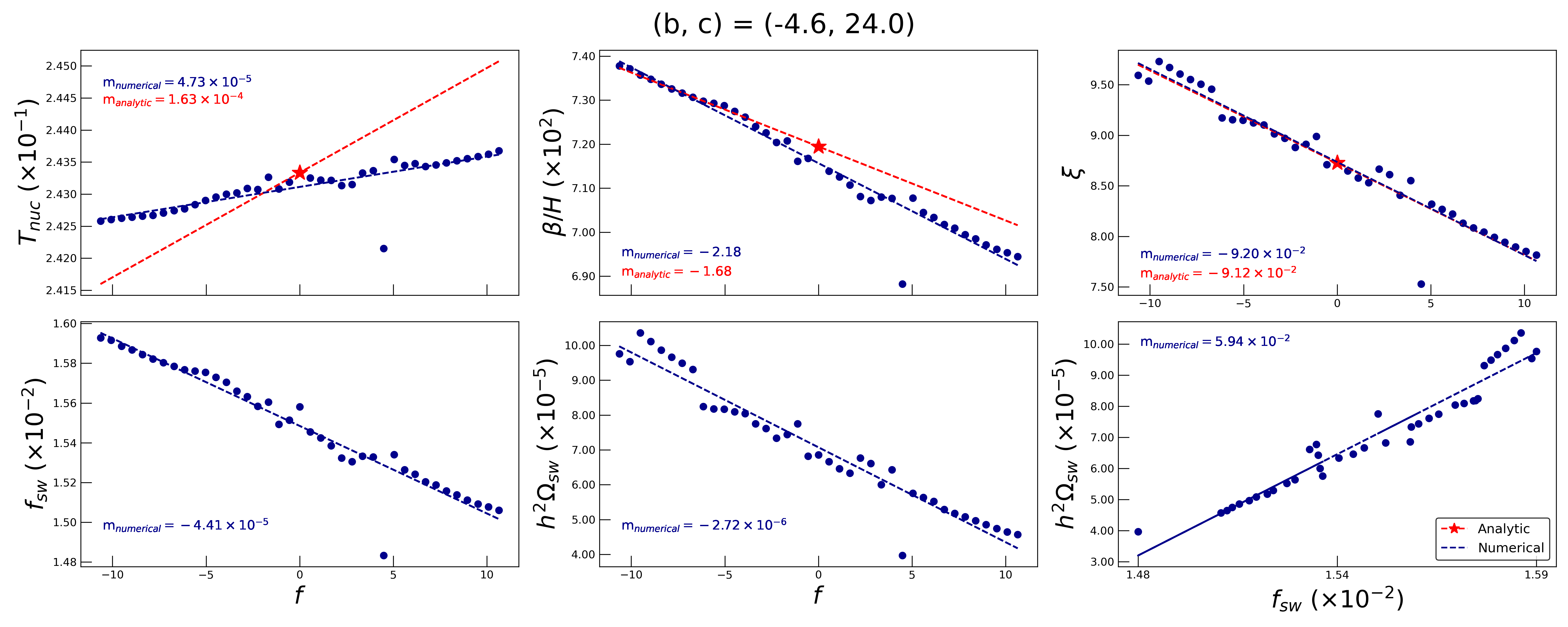} \\
     \captionsetup{justification   = RaggedRight,
             labelfont = bf}
    \caption{$T_N$ (in units of $v$, upper left panel), $\beta/H$ (upper middle panel) and $\xi$ (upper right panel), as functions of $f$, for three benchmark choices of $(b,c)$, as labeled.  Also shown are the peak values $f_{sw}$ (lower right panel) and $h^2 \Omega_{sw}$ (lower middle panel) as functions of $f$, and $h^2 \Omega_{sw}$ as a function of $f_{sw}$ (lower right panel).  The blue dots are obtained from \texttt{CosmoTransitions}, and the dashed blue line is a linear fit to these points.  The red star is a prediction from the analytic form of $S_3/T$, and the red dashed line is the analytic fit.  Note that, in each case, the blue dot corresponding to $f=0$ lies under the red star.}
    \label{fig:Benchmarks}
\end{figure} \end{widetext}

In Figure~\ref{fig:Benchmarks}, we plot the thermal parameters $T_N$ (in units of $v$, upper left panels), $\beta/H$ (upper middle panels), and $\xi$ (upper right panels) obtained from \texttt{CosmoTransitions} (blue dots) as a function of $f$ for a few benchmark choices of the parameters $(b,c)$ (labeled).   In each of these panels, the dashed blue line is a linear fit to the results obtained from \texttt{CosmoTransitions}.  The red star is the value which would be obtained at $f=0$ if one took the Euclidean action evaluated on the bounce solution to be given by Eq.~\ref{eq:S3Tapprox}.  The red dashed line is the linear extrapolation from this point given by our analytic estimates in the previous section. We also plot the peak gravitational wave frequency sourced by sound waves ($f_{sw}^{max}$) and peak amplitude ($h^2 \Omega_{sw}^{max}$) as functions of $f$ in the lower left and lower middle panels, respectively.  We take $n=1$. Finally, we plot $h^2 \Omega_{sw}^{max}$ against $f_{sw}^{max}$ (as $f$ is varied) in the lower right panel.   Note that these benchmark points are only chosen to illustrate the changes in the thermal parameters of the FOPT which result from low-temperature thermal corrections, not necessarily because of their interest as targets for current or upcoming gravitational wave observatories. Although we have shown only a few benchmark points, these results are typical of the parameter points we scanned over.

We see from this figure that, generally, the full numerical analysis reproduces our analytic estimates quite well.  Deviations between the numerical and analytic results are largest for the nucleation temperature, but both the numerical and analytic results show that the variation in the nucleation temperature with $f$ is small, as argued earlier in Section~\ref{sec:Analytical}. Note that the analytic estimate was derived without any use of the detailed form of the window function.  The fact that it matches so well with a numerical calculation using the potential in Eq.~\ref{eq:Vthermal} suggests that our  assumption that the corrections to the thermal parameters depend largely on the magnitude of the low-temperature corrections at the false vacuum is borne out. The corrections to the thermal parameters seem relatively insensitive to the details of how the low-temperature effective potential correction terms vary between the false and true vacua.  

Note also that the analytic estimate for the corrections to the thermal parameters depends on the uncorrected thermal parameters $T_N$, $\beta/H$, and $\xi$, as well as $C(T_N)$ and $\Delta V(T_N)$.  Although the detailed shape of the potential (which we took to be of quartic form) determines these thermal parameters, the corrections to the thermal parameters only depend on the shape of the potential implicitly, though the uncorrected thermal parameters.  Indeed, the analytic estimates made no use of the shape of the potential at all.  This suggests that these estimates for the impact of low-temperature thermal corrections on the thermal parameters should be valid generally, beyond the assumption of a quartic potential, provided the thin wall approximation holds.

It is not clear that all of the detailed features in the variation of the thermal parameters with $f$ obtained from \texttt{CosmoTransitions} are physical, as opposed to artifacts of the numerical computation.  But the general correspondence between the the numerical calculation and the analytic approximation over the full range of $f$ provides some confidence in the overall trends seen in the numerical calculation, independent of the sharp small-scale features.

As we see from Fig.~\ref{fig:Benchmarks}, the latent heat parameter, $\xi$, can shift by as much as $10\%$ for some of these benchmark points, while still remaining within the regime in which our analytic approximations are valid.  The correction to the fractional gravitational wave amplitude will be even larger.  We thus see that, although we have restricted ourselves to the regime in which the low-temperature corrections do not qualitatively alter the vacuum structure or the phase transition, the quantiative effect is non-negligible.  We also find that, generally, the amplitude of the gravitational wave signal decreases as $f$ increases, as the dependence of the $h^2 \Omega_{sw}$ on $\xi$ tends to dominate.


\section{Conclusions} \label{sec:Conclusions}

We have considered the impact of low-temperature thermal corrections on the parameters of first-order phase transitions.  These corrections arise from degrees of freedom whose field-dependent masses in the true, symmetry-breaking vacuum are much larger than the nucleation temperature, though in the false, symmetry-preserving vacuum they may be much smaller than the nucleation temperature. Although the form of these corrections can be very complicated, and depend in detail on the couplings of the degrees of freedom which are integrated out, we have argued that the net effect of these corrections can be encoded in a single parameter, $f$, which specifies the temperature-dependent shift in the effective potential at the false vacuum relative to the true vacuum.

We find that the nucleation temperature ($T_N$), latent heat parameter ($\xi$) and transition rate parameter parameter ($\beta/H$) all shift approximately linearly with the shift in the difference between the effective potential at the false and true vacua ($\delta \Delta V$), although with different signs.  In particular, an increase in $\Delta V$ leads to a linear decrease in $\xi$ and $\beta / H$, and a generally small increase in $T_N$.  We argue that these results are largely independent of the details of the tree-level potential and of the precise form of low-temperature corrections, provided the thin-wall approximation is valid.   

We illustrate these results by considering the specific case in which the thermal effective potential (absent the low-temperature corrections) is of quartic form.  Considering several benchmark cases, we show that the results we obtain from analytic estimation is largely borne out by more detailed numerical calculation with \texttt{CosmoTransitions}, using a simple parameterization of the low-temperature corrections.  The fact that the numerical calculation matches the results from analytic estimation tends to confirm the genericity of the analytic estimate, and that the effect of all low-temperature corrections can be parameterized by a single strength parameter.  We expect that these results can be generally applied to the low-temperature thermal corrections to any tree-level potential, provided they are small.

Gravitational waves can be produced by a first-order phase transition, in which case the shift to the frequency and amplitude of this signal is determined by the latent heat and transition rate parameters.  In particular, an increase in $\Delta V$ will lead to a decrease in the frequency of the gravitational wave signal.  On the other hand, the impact on the amplitude of the signal depends on the details of the model, since it depends on a competition between corrections to $\xi$ and $\beta / H$, though generally the dependence on $\xi$ dominates, leading to a decreasing gravitational wave amplutide as $f$ increases.  It is worth noting that the dependence of the gravitational wave signal arising from an FOPT on the thermal parameters of the FOPT has gone through much evolution over recent years, as more detailed numerical simulations have been used to model the generation of sound waves in the plasma.  But the rough trends which we describe here are largely determined by dimensional analysis and power counting, and are thus largely robust to further developments in this field.  Moreover, as systematic uncertainties in the generation of gravitational waves are reduced by the further study of improved simulations, the impact of low-temperature corrections will be even more significant.

We also note the impact of our results on cosmology.  In particular, if $f$ parameterizes the shift in $\Delta V$, we roughly find $\delta \xi \sim - 0.1 f~ (g_*/100)^{-1}$, where $|f| \sim {\cal O}(0.1)$ for a single degree of freedom.  Consider a dark sector FOPT at late times, when $g_* \sim {\cal O}(1)$, assuming the dark sector and SM sector are at the same temperature.  We then see that low-temperature corrections from even a single degree of freedom would shift $N_{eff}$ by an ${\cal O}(1)$ number, implying that one must understand all the degrees of freedom which might gain mass from the dark Higgs, even if they are very heavy in the true vacuum, in order to understand if the model is consistent with constraints from observational cosmology.  On the other hand, in many commonly considered models, the dark sector temperature is much lower than the SM temperature.  In this case, fermionic degrees of freedom (which yield a positive contribution to $f$) will only slightly reduce $\Delta N_{eff}$, potentially alleviating the generally tight constraints placed on dark sector FOPTs by bounds on the latent heat which can be released.

\begin{acknowledgments}
JBD and JK acknowledge the Center for Theoretical Underground Physics and Related Areas (CETUP*), the Institute for Underground Science at Sanford Underground Research Facility (SURF), and the South Dakota Science and Technology Authority for hospitality and financial support, as well as for providing a stimulating environment where part of this work was completed. We gratefully acknowledge Jack Runburg for early work in developing our code for utilizing the \texttt{CosmoTranistions} package. KC is supported in part by NASA grant \#22-FERMI22-0011. JK is supported in part by DOE grant DE-SC0010504. JBD and JOW are supported by the National Science Foundation under grant no. PHY-2412995. JBD thanks the Mitchell Institute at Texas A\&M University for its hospitality where part of this work was completed. 
\end{acknowledgments}

\appendix


\section{Variations to the Window Function}
\label{sec:window}

\begin{figure}[th]
    \centering
    \includegraphics[width=0.9\linewidth]{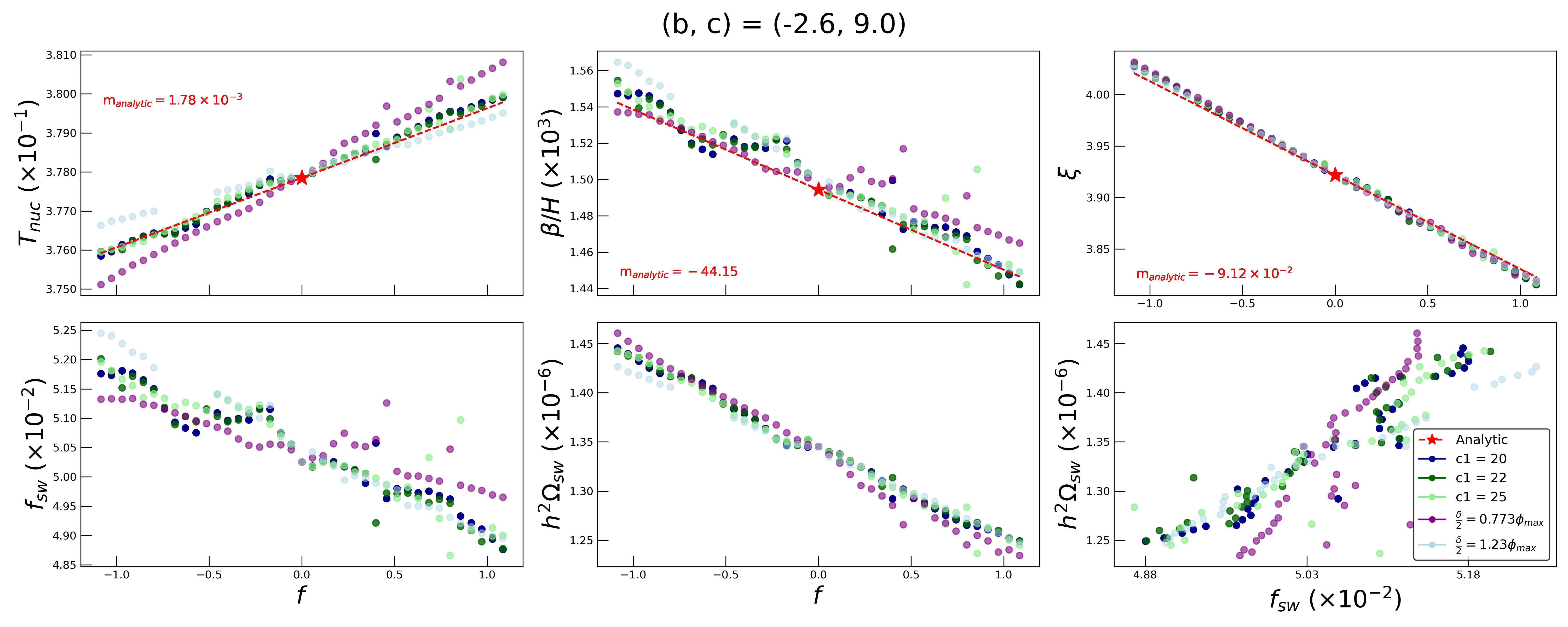}
    \captionsetup{justification   = RaggedRight,
             labelfont = bf}
    \caption{We replot the first benchmark point ($(b,c) = (-2.6,9.0)$) from Figure~\ref{fig:Benchmarks}.  
    In addition to the nominal choices of the window function parameters $c_1 =20$ and $\delta/2 = \phi_{max}$, 
    we also plot     points with $c_1 = 22, 25$ (keeping $\delta$ fixed at its nominal value), and with 
$\delta / 2 = 0.773\phi_{max}, 1.23\phi_{max}$ (keeping $c_1$ fixed at its nominal value).}
    \label{fig:VaryingWindowFunction}
\end{figure}

To illustrate how our numerical results depend on the precise choice of window function parameters $c_1$ and $\delta$, we replot our first benchmark point ($(b,c) = (-2.6,9.0)$) in Figure~\ref{fig:VaryingWindowFunction}. In addition to our nominal values of $c_1 = 20$, $\delta / 2 = \phi_{max}$, we also plot points with $c_1 = 22, 25$ (keeping $\delta$ fixed at its nominal value), and with $\delta / 2 = 0.773 \phi_{max}, 1.23 \phi_{max}$ (keeping $c_1$ fixed at its nominal value).  These values of $\delta$ were chosen to satisfy the relation  $V(\delta/2, T_C) = 0.9 V(\phi_{max},T_C)$.  Since our numerical results are well described by an analytic approximation which did not utilize the form of the window function, we expect that our results should be robust to changes in the window function parameters.  We see from Figure~\ref{fig:VaryingWindowFunction} that, as expected, our results vary only slightly under modifications of the window function.  Note in particular that $\xi$, for which the analytic estimate provided the best description of the numerical result, also shows the least variation under changes of the window function parameters.

\onecolumngrid

\twocolumngrid


\section{Bubble Solution}
\label{sec:BubbleSolution}
\vskip-0.7\baselineskip
\par

\begin{figure}[!t]
    \centering
    \includegraphics[
        width=\linewidth,
        height=5cm
    ]{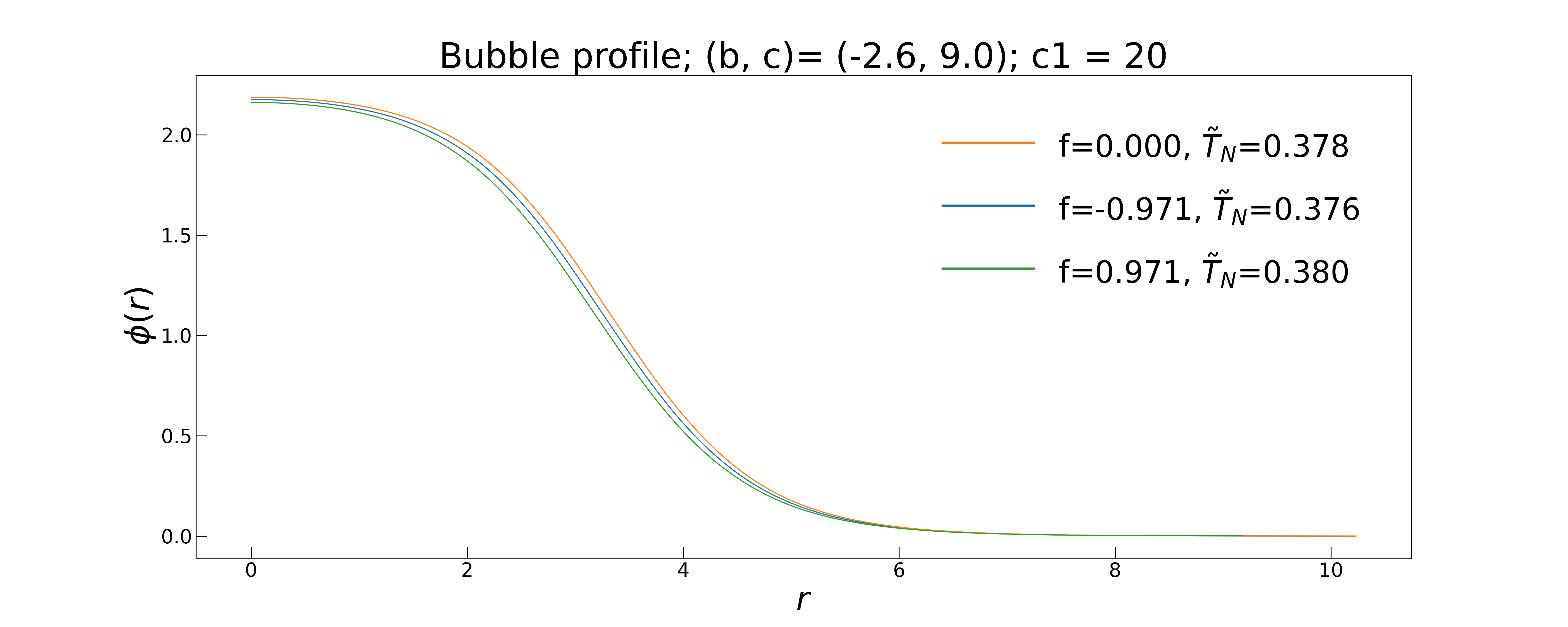}

    \captionsetup{
        justification=RaggedRight,
        singlelinecheck=false,
        labelfont=bf
    }
    \caption{%
        We plot $\phi(r)$ as a function of $r$ for the bubble
        solutions obtained by \texttt{CosmoTransitions} at the
        nucleation temperature for the thermal effective potential
        at the benchmark point $(b,c)=(-2.6,9.0)$, with $f=0$ or
        $f=\pm0.971$, as labeled.
    }
    \label{fig:BubbleSolution}
\end{figure}

In Figure~\ref{fig:BubbleSolution}, we plot the bubble solutions obtained from \texttt{CosmoTransitions} for the benchmark point $(b,c) = (-2.6,9.0)$ at the nucleation temperature, for $f=0$ and $f = \pm 0.971$.  These $f$ values lie near the extremes of the range we consider.  We see that the thin wall approximation is at least reasonable, and that the bubble solution varies smoothly as we vary $f$ within this range.  This  confirms that the dynamics of the phase transition do not change qualitatively, and that our parametric treatment of the low temperature corrections is applicable.

\bibliographystyle{apsrev4-1.bst}

\end{document}